\documentclass[twocolumn,preprintnumbers,amsmath,amssymb,floatfix,prl]{revtex4}

\usepackage{graphicx}
\usepackage{bm}

\usepackage{graphics}
\usepackage{color}

\newcommand{\LIA}{\lambda_{\rm I}^{\rm A}}
\newcommand{\LIB}{\lambda_{\rm I}^{\rm B}}
\newcommand{\LR}{\lambda_{\rm R}}
\newcommand{\LPIA}{\lambda_{\rm PIA}}

\begin{document}


\title{Graphene on transition-metal dichalcogenides: a platform for proximity spin-orbit physics and optospintronics}
\author{Martin Gmitra and Jaroslav Fabian}
\affiliation{Institute for Theoretical Physics, University of Regensburg, 93040 Regensburg, Germany\\
}

\begin{abstract}
Hybrids of graphene and two dimensional transition metal dichalcogenides (TMDC) 
have the potential to bring graphene spintronics to the next level. As we show here
by performing first-principles calculations of graphene on monolayer MoS$_2$, there
are several advantages of such hybrids over pristine graphene. First, Dirac electrons
in graphene exhibit a giant global proximity spin-orbit coupling, without compromising
the semimetallic character of the whole system at zero field. Remarkably, 
these spin-orbit effects can be very accurately described by a simple 
effective Hamiltonian. Second, the Fermi level can be tuned by a transverse electric
field to cross the MoS$_2$ conduction band, creating  a system of 
coupled massive and massles electron gases. Both charge and spin transport
in such  systems should be unique. Finally, we propose to use graphene/TMDC 
structures as a platform for optospintronics, in particular for optical spin 
injection into graphene and for studying spin transfer between TMDC and graphene. 
\end{abstract}

\maketitle

Graphene spintronics \cite{Han2014:NatNano} has relied exclusively on electrical spin injection
\cite{Tombros2007:N, Pi2010:PRL, Yang2011:PRL}. Combining
graphene with semiconducting two-dimensional TDMC \cite{Mak2010:PRL} can
open new venues for spintronics applications \cite{Zutic2004:RMP, Fabian2007:APS}. 
Indeed, TMDC are becoming
increasingly popular in optoelectronics as sensitive photodetectors \cite{Lopez-Sanchez2013:NatNanotech} or, forming lateral heterostructures 
\cite{Huang2014:NatMat,Lee2014:NatNano}, as two-dimensional solar cells \cite{Pospischil2014:NatNano}.
Important, TMDC have a sizeable
spin-orbit coupling and lack space inversion symmetry. As a result, their band
structure \cite{Kormanyos2015:2DM} allows for a valley resolved optical spin excitation
by circularly polarized light \cite{Xiao2012:PRL,Mak2012:NatNano,Zeng2012:NatNano}.
TMDC can thus facilitate optical spin injection into graphene, in hybrid structures. 

Efficient growth of MoS$_2$ on graphene has already been demonstrated 
\cite{Lin2014:ACS, Lin2014:APL, Azizi2015:ACS}. It was reported that graphene 
on MoS$_2$ is ultraflat, having large mean free paths \cite{Lu2014:PRL}; 
angle-resolved photoemission found an intact Dirac point 
but a strong hybridization elsewhere in the $\pi$ system \cite{Diaz2015:NL}.
Technological potentials for these hybrid structures are already
being discussed \cite{Kumar2015:MT}, mainly
as a basis for nonvolatile memory \cite{Bertolazzi2013:ACSNano}, 
sensitive photodetection \cite{Zhang2014:SREP},
and gate-tunable persistent photoconductivity \cite{Roy2013:NatNanotech}.
Recently, the spin Hall effect in graphene on few-layer  WS$_2$ was observed 
at room temperature \cite{Avsar2014:NatComm}.

In this paper we establish by first-principles calculations fundamental electronic properties and the 
spin-orbit fine structure of the graphene Dirac bands for graphene on monolayer MoS$_2$, and introduce an effective spin-orbit Hamiltonian which explains
the proximity induced spin splittings of the Dirac states. We show that
the induced spin-orbit coupling is giant, being 20 times more than in pristine graphene. 
We also discuss the field effect on the band offsets of the two materials. 
Finally, we present possible experimental schemes to perform optical spin injection into 
graphene and study spin tunneling from TMDC through graphene.

\begin{figure}[h!]
 \includegraphics[width=0.99\columnwidth]{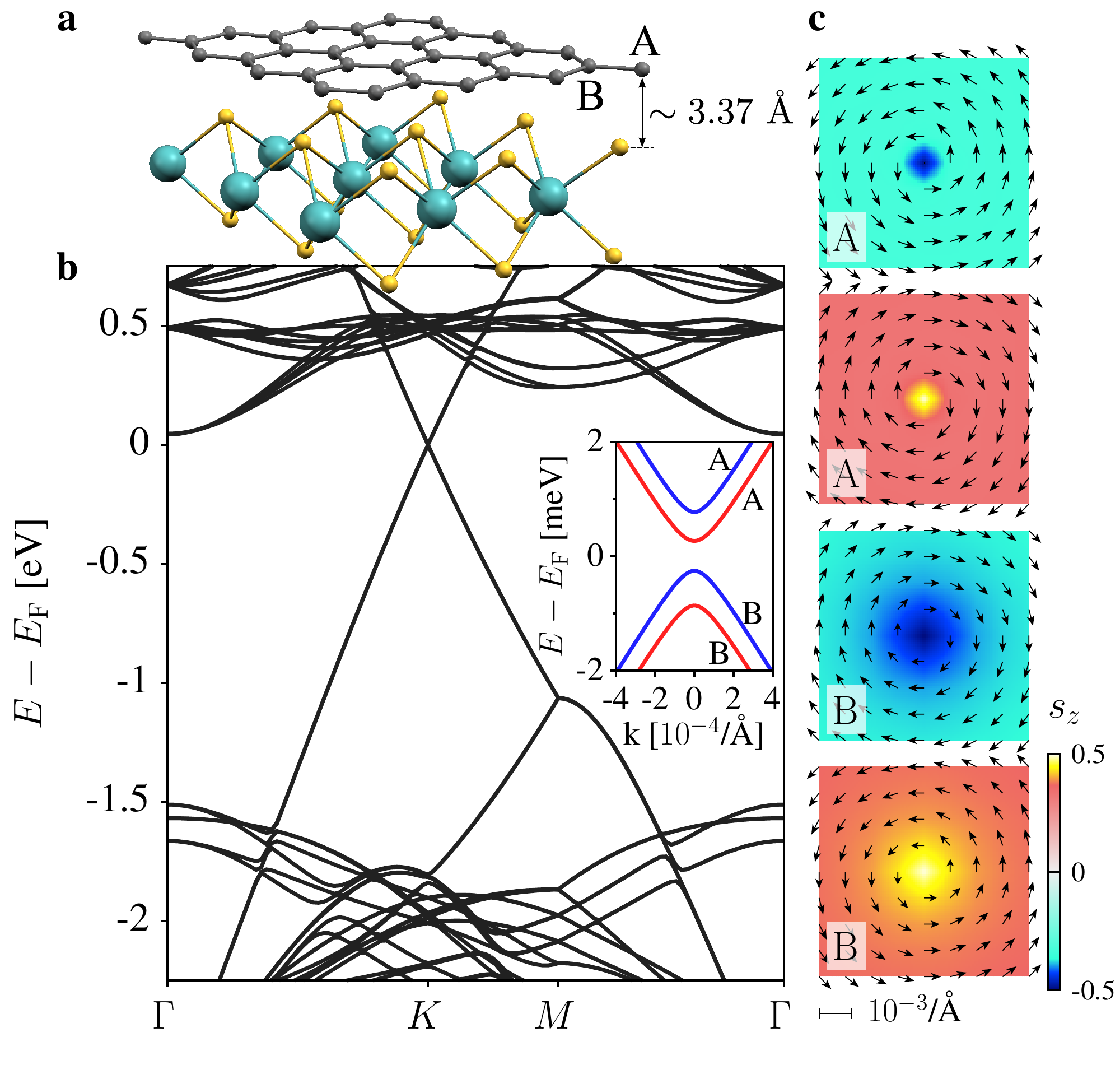}
 \caption{Calculated electronic and spin properties of graphene on two-dimensional MoS$_2$.
(a)~The supercell used in first-principles calculations.
(b)~Calculated band structure along high symmetry lines. The inset is a zoom to the fine structure of the
low energy
bands at the Fermi level, around the Dirac point. Bands with positive (negative) $z$ component of the spin are shown in red (blue). The sublattice character ($A$ and $B$) is also indicated.
(c)~Spin textures for the four bands of the inset in b).
 }\label{Fig:bands}
\end{figure}

\paragraph{First-principles results.}
To establish the electronic and spin properties of graphene on MoS$_2$ we used first-principles methods
based on density-functional theory \cite{Hohenberg1964:PR}, see Methods. 
To reduce structural strain we constructed a large supercell of 59 atoms, comprising a $3\times 3$ supercell of MoS$_2$ and a $4\times 4$ supercell of graphene, with the residual lattice mismatch of 1.4\%. 
The relaxed interlayer distance between graphene and MoS$_2$ is 3.37~\AA, see
Fig.~\ref{Fig:bands}a).
In this supercell the $K$ point of MoS$_2$ is mapped to the $\Gamma$ point in the 
reduced Brillouin zone. The calculated electronic band structure is shown in Fig.~\ref{Fig:bands}b).
The Dirac cones of graphene are nicely preserved, with the projected Dirac point (which is also the 
Fermi level) being  slightly below the conduction band edge of MoS$_2$.
The closeness of the Dirac point to the conduction band of MoS$_2$ enhances screening, 
which can substantially increase the mean free path in the graphene layer, as recently shown 
experimentally \cite{Lu2014:PRL}.
%

\begin{figure}[h!]
 \includegraphics[width=0.99\columnwidth]{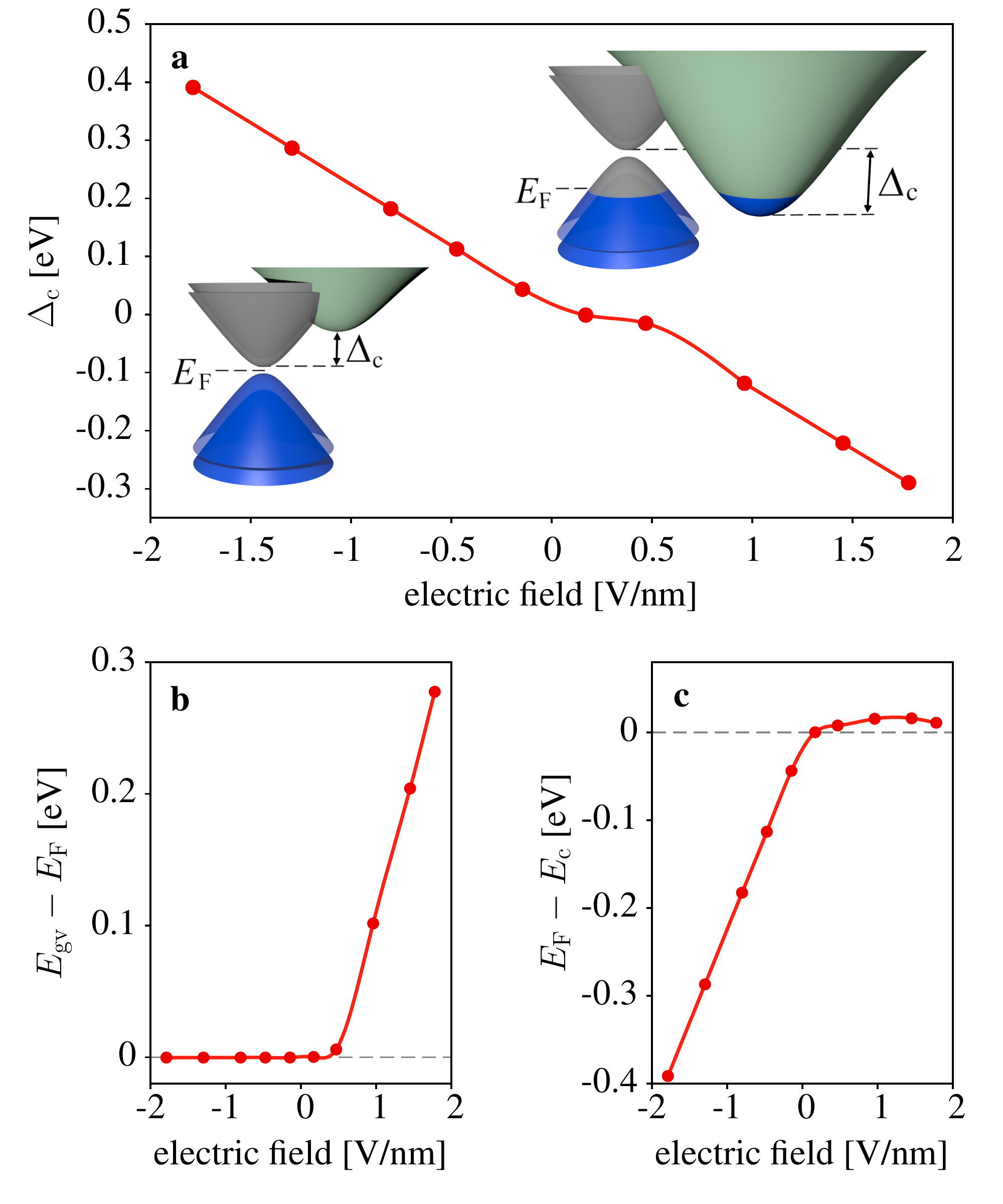}
 \caption{Field effect on graphene/MoS$_2$.
(a),~Calculated offset $\Delta_{\rm c}$ from the conduction band minimum
of MoS$_2$ to graphene, as a function of an applied transverse electric field.
At positive fields electrons are transferred from graphene to MoS$_2$, 
establishing a massless-massive electrons bilayer.
(b)~Net Fermi energy for graphene, as the difference of the valence band 
maximum $E_{\rm gv}$ of graphene and the system Fermi level, $E_{\rm F}$.
At positive fields the valence band of graphene becomes populated.
(c)~Net Fermi energy for MoS$_2$, as the difference between
the system Fermi level, $E_{\rm F}$, and the conduction band 
minimum, $E_{\rm c}$ of MoS$_2$. The conduction band
of MoS$_2$ becomes populated at positive field, reflecting
the population of holes in graphene in (b) as the whole system is neutral.
}\label{Fig:cbm}
\end{figure}

The band offsets between graphene and MoS$_2$ can be controlled by an external 
electric field applied transverse to the layers. This is demonstrated by our first-principles calculations 
 in  Fig.~\ref{Fig:cbm}a), where we present $\Delta_{\rm c}$, the difference between the conduction 
band minima of  MoS$_2$ and graphene. At negative fields (pointing towards MoS$_2$) the offset
increases, leaving both layers neutral. However, positive fields shift the Dirac point above the 
conduction band minimum of MoS$_2$ and populate graphene with holes and MoS$_2$ with 
electrons. The Fermi level crosses both the valence band of graphene and the conduction band of MoS$_2$, see  Fig.~\ref{Fig:cbm}b) and c). This field effect can establish a unique system in which massless Dirac
electrons are coupled with a conventional 2d electron gas \cite{Scharf2012:PRB}.

We now zoom in on the Dirac point at $K$ to see how the electronic spectrum of graphene deforms
in the presence of MoS$_2$. This fine structure is shown in the inset to Fig.~\ref{Fig:bands}b). 
There are two important effects: First, an orbital band gap opens, 
due to the breaking of the graphene pseudospin symmetry. On average, 
atoms $A$ and $B$ in the graphene supercell see a different environment
coming from the MoS$_2$ layer. This orbital
gap is there even in the absence of spin-orbit coupling. It arises from the 
effective staggered potential induced by the pseudospin symmetry breaking. 
Second, spin-orbit coupling combined with the broken space inversion 
symmetry lifts the spin degeneracy of the Dirac valence and conduction
bands and leads to the appearance of four distinct bands. This 
splitting is on the meV scale, which is giant when compared to the 
24 $\mu$eV spin-orbit splitting in pristine graphene \cite{Gmitra2009:PRB}. 
The inset also shows the orbital character of the bands at $K$:
while the valence states are formed at the $B$ 
sublattice, the conduction states live on $A$. The same orbital ordering is at $K'$.

Another important characteric of the Dirac states is their spin texture. 
This is plotted in Fig.~\ref{Fig:bands}c) for the four bands from the inset of  Fig.~\ref{Fig:bands}b). 
Directly at $K$ the spins are pointing out of the graphene plane, alternating up 
and down. Increasing the momentum away from $K$, the spins acquire a
 winding in-plane component, either clockwise or counterclockwise, suggestive
of the strong Rashba effect. At $K'$ the spins are reversed.

\paragraph{Effective Hamiltonian.}
Can we understand these proximity-induced changes in graphene's band structure from an effective model? 
The answer is not obvious since not only sublattices 
$A$ and $B$ differ, but even the sites that belong to the same sublattice 
see different local environments in the supercell. Surprisingly, 
an effective symmetry-based Hamiltonian with graphene orbitals 
in the presence of pseudospin inversion symmetry breaking gives a 
remarkably good description. The model builds on the orbital Hamiltonian for pristine graphene which, 
close to $K$($K'$) points, is
\begin{equation}
 {\cal H}_0=\hbar v_{\rm F}(\kappa\sigma_x k_x+\sigma_y k_y).
\end{equation} 
Here $v_{\rm F}$ is the Fermi velocity of graphene, $k_x$ and $k_y$ are the Cartesian components
of the electron wave vector measured from $K(K')$, parameter $\kappa = 1\,(-1)$ 
for $K\,(K')$, and $\sigma_x$ and $\sigma_y$ are the pseudospin
Pauli matrices acting on the two-dimensional vector space formed by the 
two triangular sublattices of graphene. 
Hamiltonian ${\cal H}_0$ describes gapless Dirac states with the conical 
dispersion $\varepsilon_0=\nu\hbar v_{\rm F}|\bm{k}|$ near the Dirac points; 
$\nu=1(-1)$ for the conduction (valence) band.

The staggered potential describing the effective orbital energy difference on $A$ and $B$ sublattices
of graphene on MoS$_2$ enters via the Hamiltonian,
\begin{equation}
{\cal H}_{\rm \Delta}=\Delta\,\sigma_z s_0, 
\end{equation}
where $\sigma_z$ is the pseudospin Pauli matrix and $s_0$ is the unit spin matrix;
$\Delta$  is the proximity induced gap of the Dirac spectrum. Another consequence
of the pseudospin inversion asymmetry is the sublattice-resolved 
intrinsic spin-orbit coupling. Indeed, the intrinsic coupling acts solely on a
given sublattice: it is a next-nearest neighbor hopping \cite{Konschuh2010:PRB}. We describe it in our model
with parameters $\LIA$ and $\LIB$ for sublattices $A$ and $B$, respectively. 
The corresponding proximity induced spin-orbit coupling Hamiltonian 
close to $K(K')$,
\begin{equation}
{\cal H}_{\rm SO}= \LIA[(\sigma_z+\sigma_0)/2] \kappa s_z+ 
\LIB[(\sigma_z-\sigma_0)/2]\kappa s_z,
\end{equation}
is a generalization of the McClure-Yafet Hamiltonian for graphene 
\cite{j._w._mcclure_theory_1962, Han2014:NatNano}. We denote
by $s_z$ the spin Pauli matrix, while by $\sigma_0$ the unit matrix acting
on the pseudospin (sublattice) space. If $\LIA = \LIB$, the main effect of the intrinsic 
spin-orbit coupling is to enhance the anticrossing of the pristine graphene 
Dirac cones \cite{Gmitra2009:PRB}, leaving the spin degeneracy intact. However, 
if $\LIA \ne \LIB$, as in our case, the spin degeneracy gets lifted by this intrinsic term already,
reflecting the loss of space inversion symmetry.

Placing graphene on MoS$_2$ also breaks the lateral mirror symmetry,
giving rise to the Rashba type spin-orbit coupling \cite{Konschuh2010:PRB},  
\begin{equation}
{\cal H}_{\rm R}=\LR(\kappa\sigma_x s_y-\sigma_y s_x), 
\end{equation}
where $\LR$ is the Rashba parameter and  $s_x$, $s_y$ are the spin Pauli matrices. In the  hopping
language, the Rashba coupling is the nearest-neighbor spin-flip hopping, contributing further to the spin splitting
of the bands, and defining the spin quantization axis for each Bloch state, away from the time reversal
points $\Gamma$ and $M$.

Hamiltonian ${\cal H}_0 + {\cal H}_{\rm \Delta} + {\cal H}_{\rm SO} + {\cal H}_{\rm R}$ fully describes
graphene's bands at $K(K')$. Its eigenenergies are
\begin{eqnarray}\label{Eq:spectrum}
\varepsilon_{\nu\mu} = \frac{1+\nu\mu}{2}\left[\nu\Delta+\frac{1+\nu}{2} \LIA+\frac{1-\nu}{2} \LIB\right] - \nonumber\\
\frac{1-\nu\mu}{4}\left[\LIA+\LIB-\nu\sqrt{(2\Delta-\LIA+\LIB)^2 + 16\LR^2}\right],
\end{eqnarray}
where $\mu=1(-1)$ for spin up (down) branches. The expectation values of the spin along
$z$ for the corresponding states are given by
\begin{widetext}
\begin{equation}\label{Eq:spin}
\langle s_z\rangle_{\nu\mu} = \frac{\mu\kappa\hbar}{2}\left[\frac{1+\nu\mu}{2}+
\frac{1-\nu\mu}{2}
\frac{2\Delta-\LIA-\LIB}{\sqrt{(2\Delta -\LIA+\LIB)^2 + 16\LR^2}}\right].
\end{equation}
\end{widetext}

%
\begin{figure}[h!]
 \includegraphics[width=0.99\columnwidth]{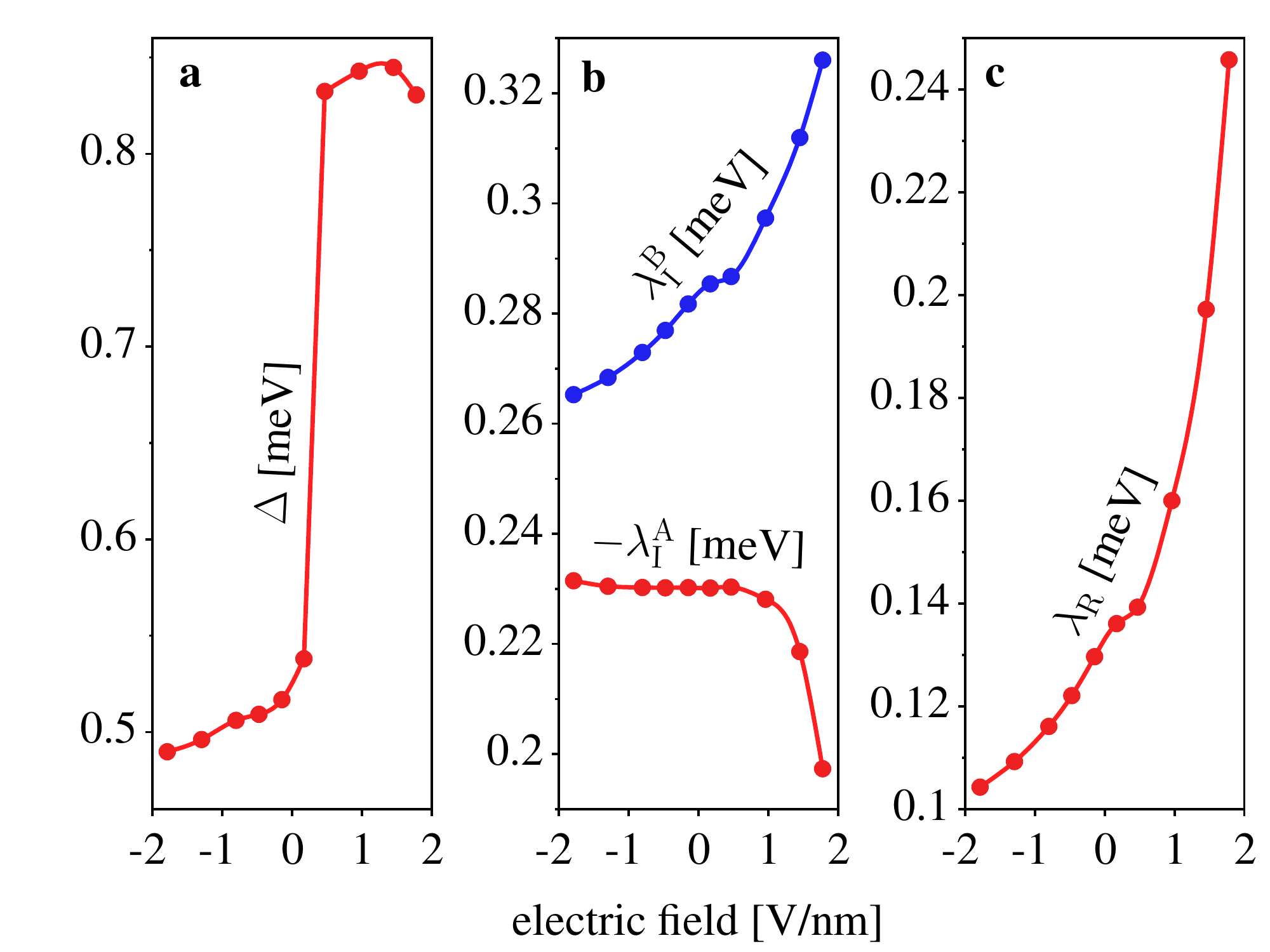}
 \caption{Calculated effective Hamiltonian parameters at the K point: 
(a)~hybridization gap $\Delta$, 
(b)~sublattice resolved intrinsic spin-orbit coupling $\LIA$ and $\LIB$ (they 
have opposite signs), and 
(c)~Rashba parameter $\LR$, as  functions of the applied transverse electric field.
 }\label{Fig:soc}
\end{figure}

Using the formulas for the eigenenergies, Eq.(\ref{Eq:spectrum}), and for the
spin expectation values, Eq.(\ref{Eq:spin}), we can algebraically extract the orbital band gap $\Delta$
and the three spin-orbit parameters $\LIA$, $\LIB$, and $\LR$ by comparing to our first-principles
data for the fine structure at $K$, see the inset to Fig.~\ref{Fig:bands}b). 
The extracted parameters are shown in Fig.~\ref{Fig:soc} as a function of the applied transverse electric
field. The orbital proximity gap $\Delta$ is about 0.5 meV in zero field. In fields greater than
0.5~V/nm,  the gap exhibits a steep increase, which is related to the 
transfer of the electronic charge from graphene to MoS$_2$, see Fig.~\ref{Fig:cbm}. 
The proximity spin-orbit parameters in Fig.~\ref{Fig:soc}b) and c)  are about 0.2 meV, which is 20 times more than in pristine
graphene \cite{Gmitra2009:PRB}. Similar giant values of spin-orbit coupling in graphene are induced
by hydrogen adatoms \cite{neto_impurity-induced_2009, Gmitra2013:PRL, balakrishnan_colossal_2013}, 
and even more by
fluorine \cite{irmer_spin-orbit_2015}, though the mechanisms are different. Unlike in the adatom
cases in which the induced spin-orbit coupling is only local, in our case the
giant coupling is global. While the intrinsic parameters $\LIA$ and $\LIB$ change rather moderately
with applying the electric field, the Rashba parameter $\LR$, see Fig.~\ref{Fig:soc}c), more than 
doubles in increasing the field from -2 to 2 V/nm. 

What is the origin of the induced giant spin-orbit coupling in graphene on MoS$_2$? 
We trace the enhancement to the hybridization of the
carbon orbitals with the $d$-orbitals of Mo. We find 
only 0.3\% of $d$-orbitals at the $K$ point by analyzing the calculated density of states. But when 
we turn the spin-orbit coupling on Mo atoms in the supercell off,  the orbital 
gap in zero field remains almost unchanged ($\Delta = 0.506$~meV), 
while the spin-orbit parameters drop to their pristine graphene 
values $\LIA=24$~${\rm \mu eV}$, $\LIA=23$~${\rm \mu eV}$ 
and $\LR=10$~${\rm \mu eV}$, which are, curiosly, also determined by $d$ orbitals, but from
carbon atoms \cite{Gmitra2009:PRB}. 

Away from $K(K')$, the spin splittings depend on the momentum. In order to describe our first-principles
data,  we add the PIA (pseudospin inversion asymmetry) spin-orbit coupling 
term \cite{Gmitra2013:PRL} which, like the intrinsic coupling, represents the next nearest 
neighbor hopping, but with a spin flip. The full model Hamiltonian describes the data perfectly, 
see Supplementary information.

Although our effective model should capture the basic physics of graphene on TMDC, the extracted 
parameters are for the specific supercell of graphene on MoS$_2$. Certainly,
taking an even larger cell that could further reduce strain, or twisting the two layers as could happen
in experiments, would lead to a different set of parameters, although the orders of magnitudes would
likely stay. In a macroscopic experimental structure we expect Moire patterns which would transform our Hamiltonian 
into a Hamiltonian density, with an orbital gap and spin-orbit fields, perhaps even averaging some of 
the parameters (such as $\Delta$ and the difference between $\LIA$ and $\LIB$)  to zero. 
Our extraxted parameters can then be viewed as effective standard deviations of the spatial variations, 
suitable as input for charge and spin transport model calculations for such samples. We also expect that graphene on TMDC could produce superlattice features as in graphene on hBN \cite{Gorbachev2014:S}. 

%
\begin{figure}[h!]
 \begin{tabular}{l}
  \textbf{a}\\
  \includegraphics[width=0.85\columnwidth]{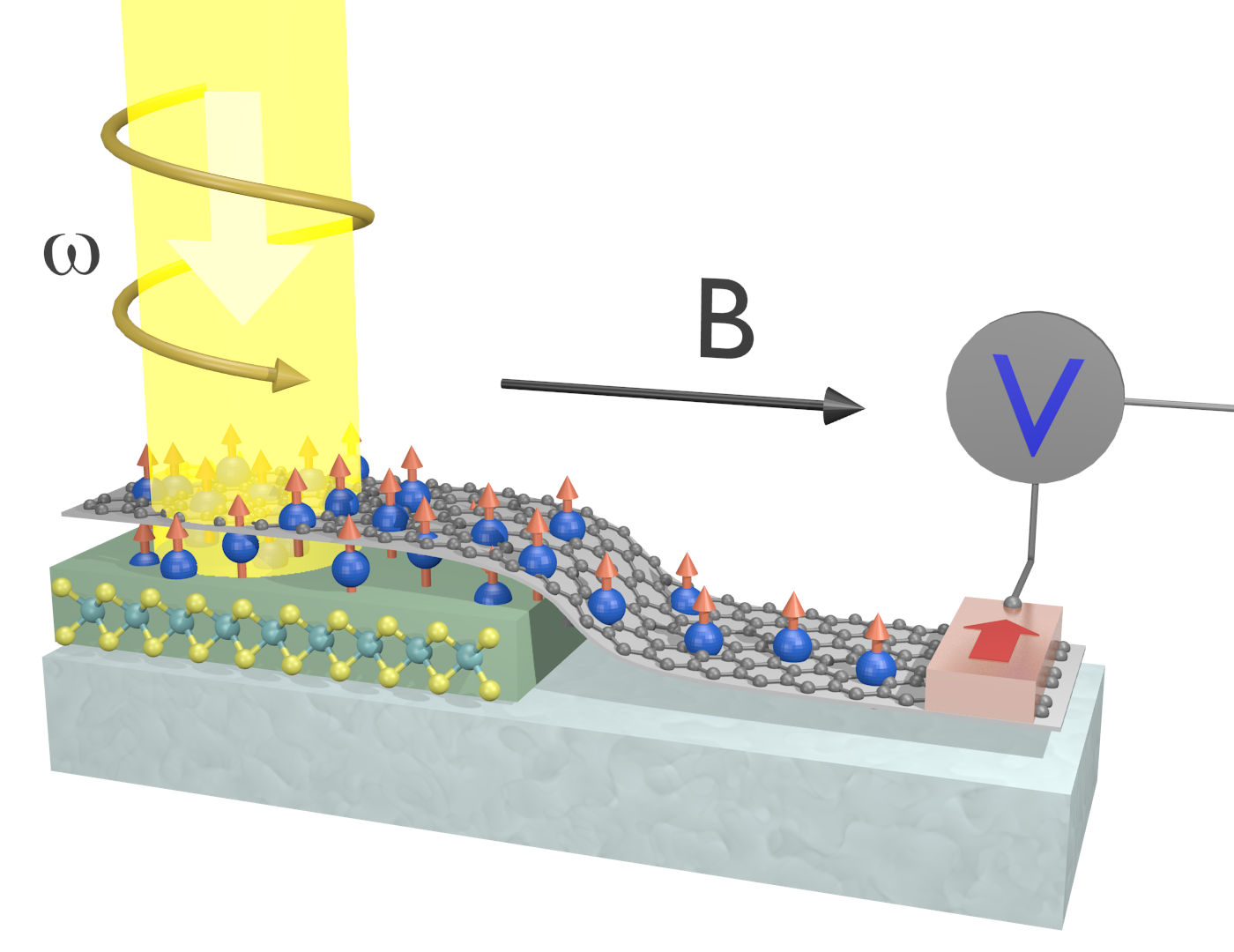}\\
  \textbf{b}\\
  \includegraphics[width=0.85\columnwidth]{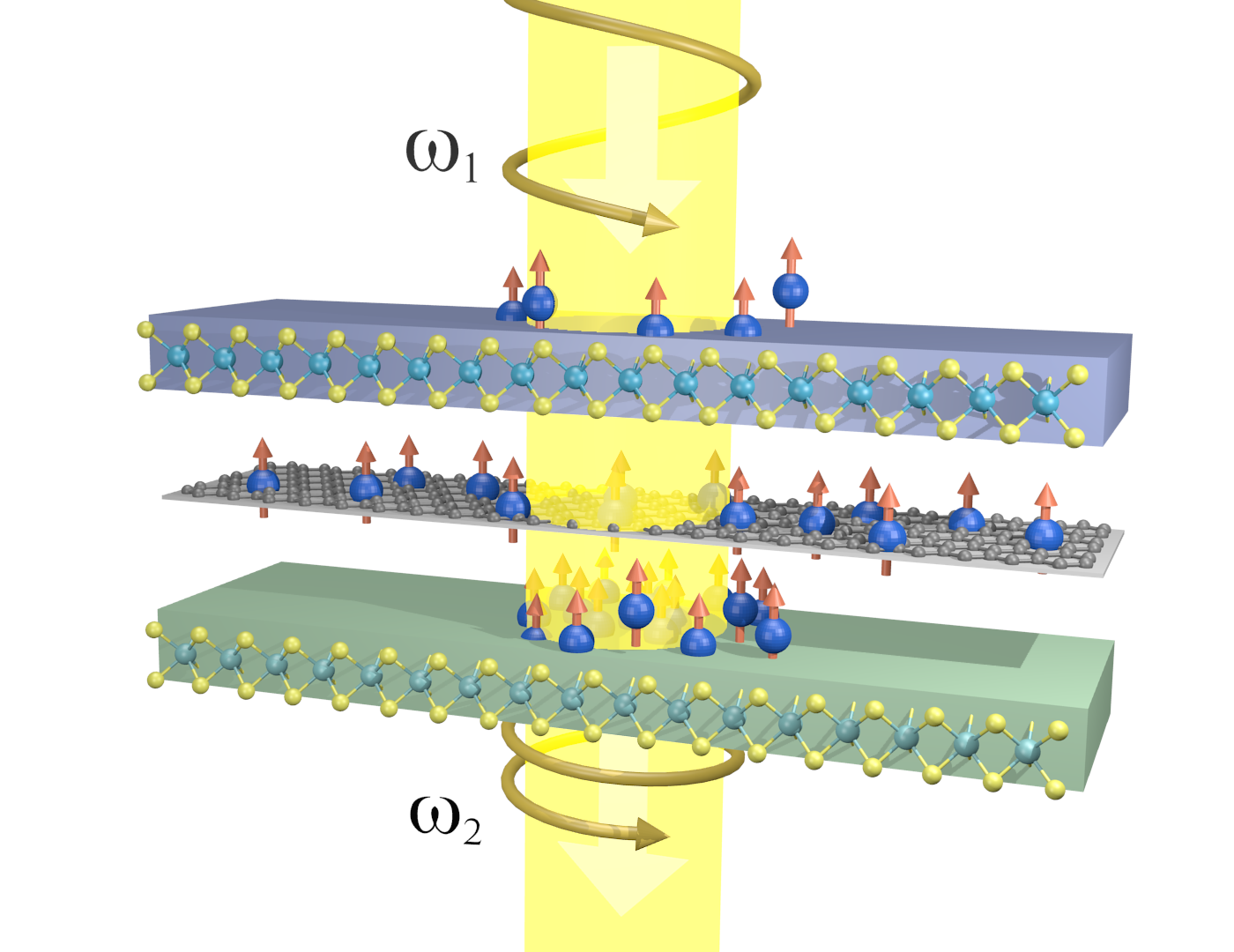}
 \end{tabular}
 \caption{Optospintronic schemes for graphene-TMDC hybrids.
(a)~Optical spin injection into graphene, facilitated by the semiconducting TMDC. 
A circularly polarized light excites spin-polarized electrons in the semiconductor.
The spin is transferred to graphene where it can be detected as a Hanle signal by the ferromagnetic
electrode.
(b)~Spin transfer between two different TMDCs, encapsulating graphene. Circularly
polarized light tuned to the band gap of the top material excites electron spins which can tunnel
to the lower material, exhibiting a circular luminescence peaked at its band gap frequency. 
 }\label{Fig:devices}
\end{figure}

\paragraph {Optospintronics.} We propose graphene-TMDC hybrids, such as the one studied above based on MoS$_2$,
 as an ideal platform for optospintronics. In 
Fig.~\ref{Fig:devices}a) we give an optical spin injection scheme into graphene.  A circularly polarized
light, tuned to the band gap of TMDC, excites electron spins by optical orientation \cite{meier_optical_1984, Zutic2004:RMP}. In effect, the light
produces spin-polarized excitons which dissociate into spin-polarized electrons and holes. 
As in the recent optical experiment \cite{Roy2013:NatNanotech}, we expect that electrons will be transfered to graphene, 
leaving holes behind in TMDC, although in which way electrons and holes split may depend on the
TMDC material as well as on gating. The spin-polarized electrons (or holes) diffuse in graphene. One can detect this
spin accumulation either optically, by observing a circular polarization of the photoluminescence \cite{meier_optical_1984}  elsewhere in graphene on TMDC, or electrically. The latter is illustrated in  Fig.~\ref{Fig:devices}a): a ferromagnetic electrode on top of graphene detects the presence of the spin accumulation in graphene \cite{Zutic2004:RMP, Fabian2007:APS}.  
Spin precession in graphene can be observed as the Hanle signal (which is not possible
to see in the spin-valley coupled TMDC \cite{Sallen2012:PRB}), by applying an external magnetic 
field transverse to the injected spin, providing Larmor precession \cite{Fabian2007:APS}. 

Spin transport {\it per se} in graphene-TMDC bilayers should be fascinating. The presence of the 
giant, effectivelly uniform spin-orbit fields should give large spin Hall signals, even greater than in hydrogenated
graphene \cite{balakrishnan_colossal_2013}. Most important, as our calculations show, the spin, like charge, properties of these structures are expected to be highly field tunable. The fascinating prospect of realizing the massive-massless electron gas coupling of the two electron gases, if the Fermi level is positioned in both
band structures, calls for new theories of spin transport and spin relaxation in such hybrid systems.

To demonstrate spin tunneling from a TMDC through graphene one could use a sandwich structure, as pictured in  Fig. ~\ref{Fig:devices}b). The two semiconductors have different band gaps, allowing to discern the photoluminesence signals from the top and bottom layers. If the spin pumping light is tuned
to the band gap of the top layer, the spin-polarized carriers would be excited there and tunnel through 
graphene to the bottom layer, in which they would recombine and emit circularly-polarized light with 
the frequency characteristics of the bottom material. One can envision influencing the signal with 
a transverse magnetic field, allowing for a Hanle effect 
Another possibility is to measure the accumulated spin in the bottom layer using the
magneto-optical Kerr effect, as in recent experiments on monolayer MoS$_2$
\cite{Korn2014:P}.

\paragraph{Conclusion.} We have established by first-principles calculations a strong effect of MoS$_2$ on the spin properties
of graphene, predicting a giant and field-tunable proximity spin-orbit coupling for Dirac electrons. We have introduced an 
effective spin-orbit Hamiltonian to describe the electronic states around the Fermi level, fitting perfectly
the first-principles data. We have also showed that gating can tune the band offsets of the two layers, allowing
to realize the unique system of coupled massless and massive electron gases. 
Finally, we have proposed to use graphene on TMDC as a platform for optospintronics
with graphene-based two-dimensional materials structures.

\begin{acknowledgments} We thank T. Korn, C. Sch\"uller, C. Stampfer, B. Beschoten, T. M\"uller, and B.  
\"Ozyilmaz for useful descussions and hints regarding possible experimental realizations of optospintronics with graphene-TMDC structures.
This work was supported by DFG SFB~689, GRK~1570, and by the EU~Seventh~Framework~Programme under Grant~Agreement~No.~604391~Graphene~Flagship.
\end{acknowledgments}

\section*{Supplementary information}

\subsection{Methods}

Structural relaxation and electronic structure
calculations were performed with {Quantum ESPRESSO} \cite{Giannozzi2009:JPCM}, using norm conserving pseudopotentials with kinetic energy cutoff of 60~Ry for wavefunctions. For the exchange-correlation potential we used
generalized gradient approximation \cite{Perdew1996:PRL}. 
The supercell containing a  $3\times 3$ supercell of MoS$_2$ and 
$4\times 4$ supercell of graphene was embeded in a slab geometry with vacuum of 
about 13~\AA,  with a dipole correction \cite{Bengtsson1999:PRB}, which is crucial
to get accurate band offsets between the Dirac point and the conduction band minimum of
MoS$_2$. The resulting structure has a lattice mismatch of 2.8\% which we split equally 
between graphene and MoS$_2$ by compressing graphene and  
stretching MoS$_2$ by 1.4\%. The supercell has 59 atoms.
The reduced Brillouin zone was sampled with $12\times 12$
k points. Atomic positions were relaxed using the quasi-newton algorithm
based on the trust radius procedure including the van der Waals interaction which was treated 
within a semiempirical approach \cite{Grimme2006:JCC,Barone2009:JCC}.
The calculated work function on the graphene side is 4.12~eV, while on MoS$_2$  it is 4.41~eV.

\subsection{Spin eigenstates}

The normalized eigenstates $\psi_{\nu\mu}$ at the $K$ point of the Hamiltonian discussed in the manuscript
in the basis $\sigma \otimes s = | A \uparrow, A \downarrow, B \uparrow, B \downarrow \rangle$ read
\begin{equation}
\begin{array}{cccccc}
\psi_{+-}=& |0,& i\alpha_+ Q_+,& 4\LR\alpha_+,& 0\rangle & \simeq |A\downarrow\rangle,\\
\psi_{++}=& |1,& 0,            & 0,           & 0\rangle & = |A\uparrow\rangle,\\
\psi_{--}=& |0,& 0,            & 0,           & 1\rangle & = |B\downarrow\rangle,\\
\psi_{-+}=& |0,& i\alpha_- Q_-,& 4\LR\alpha_-,& 0\rangle & \simeq |B\uparrow\rangle,
\end{array}
\end{equation}
where $\alpha_\pm = 1/\sqrt{16\LR^2+Q_\pm^2}$ and $Q_\pm=2\Delta+\LIA+\LIB\pm\sqrt{(2\Delta+\LIA+\LIB)^2+16\LR^2}$.
The eigenvectors are ordered with increasing energy (using the extracted parameters):
$\varepsilon_{+-}>\varepsilon_{++}>\varepsilon_{--}>\varepsilon_{-+}$.
Analyzing the above eigenvectors we see that the valence bands are localized on sublattice
$B$, while the conduction bands on sublattice $A$. The $z$ component of the spin alternates from band to band.
This behavior matches the first-principles results, see inset to Fig.~\ref{Fig:bands}b) in the manuscript.
The top valence $\varphi_{--}$ and bottom conduction
$\varphi_{++}$ states are pure pseudospin and spin states. 
On the other hand, spin-orbit coupling mixes spin and pseudospin of
the outermost states. 
The eigenstates at $K'$  have the same form, but opposite spins.

\subsection{PIA coupling: spin splitting away from $K(K')$}

To describe the calculated spin splittings away from $K$,  we employ the PIA (short for pseudospin inversion asymmetry)
spin-orbit term, introduced to study the effects of spin-orbit coupling in graphene due to hydrogen adatoms \cite{Gmitra2013:PRL}:
\begin{equation}
{\cal H}_{\rm PIA} = (\LPIA^+ \sigma_z + \LPIA^- \sigma_0) (k_x s_y - k_y s_x)\, .
\end{equation}
Here $\LPIA^+$ and $\LPIA^-$ are the spin-orbit parameters representing the average, $\LPIA^+$, 
and differential, $\LPIA^-$, PIA coupling between the $A$ and $B$ sublattices. Like intrinsic spin-orbit 
coupling, PIA can be also represented by next-nearest-neighbor (same sublattice) hopping, but with a spin flip. The PIA terms turn the spin quantization axes of the electron states towards the graphene plane and add to the Rashba term for momenta away from the $K$ point.

\begin{figure}[h!]
 \includegraphics[width=0.99\columnwidth]{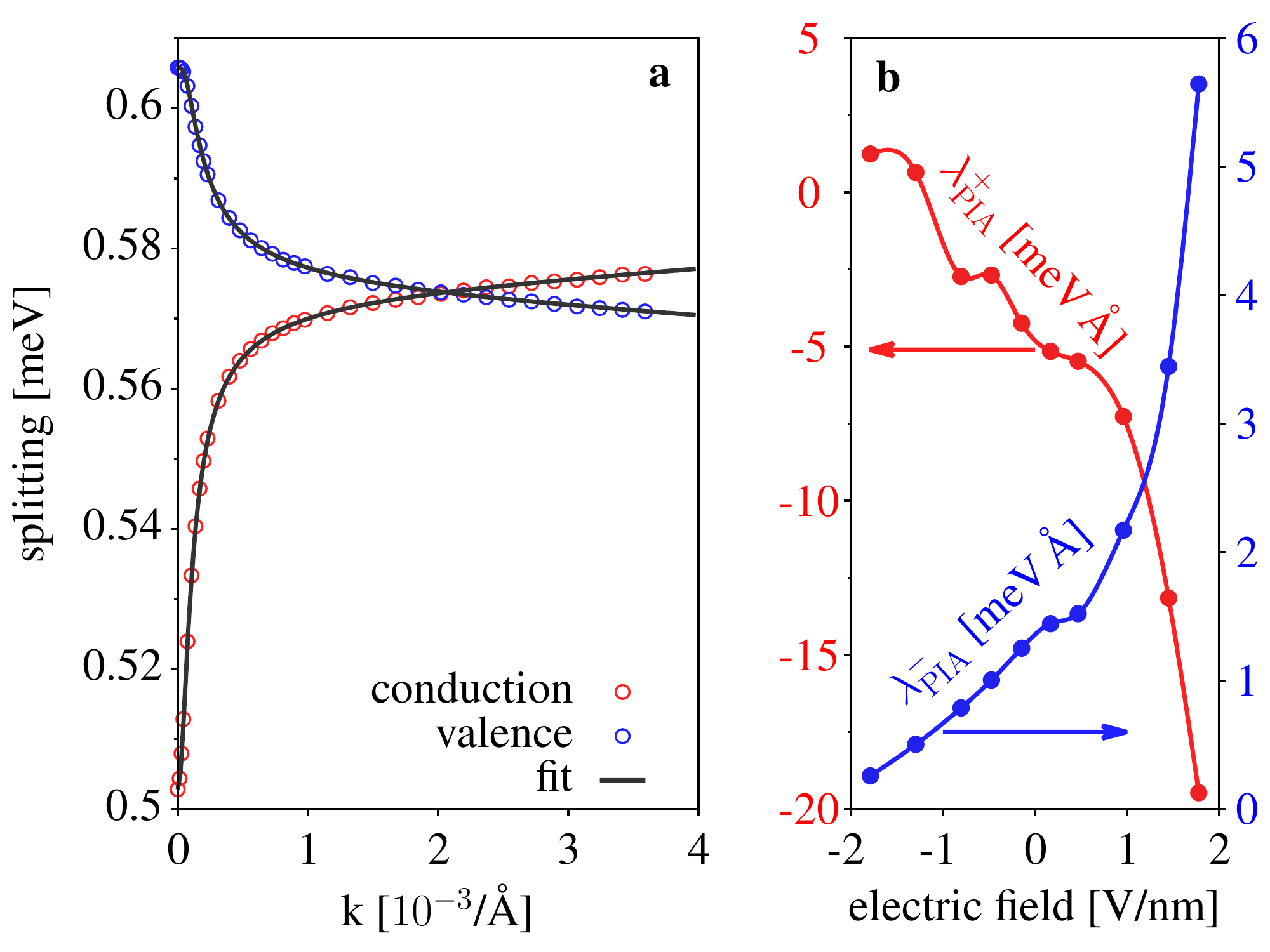}
 \caption{Spin splitting away from $K$: emergence of PIA spin-orbit coupling in graphene on MoS$_2$.
(a)~Spin splitting of conduction and valence bands from $K$ ($k=0$) in the direction towards $\Gamma$,
at zero electric field.  Solid lines are model fits, symbols are first-principles results. 
(b)~Fitted PIA parameters at different transverse electric fields. 
 }\label{Fig:PIA}
\end{figure}

\begin{figure}[h!]
 \includegraphics[width=0.99\columnwidth]{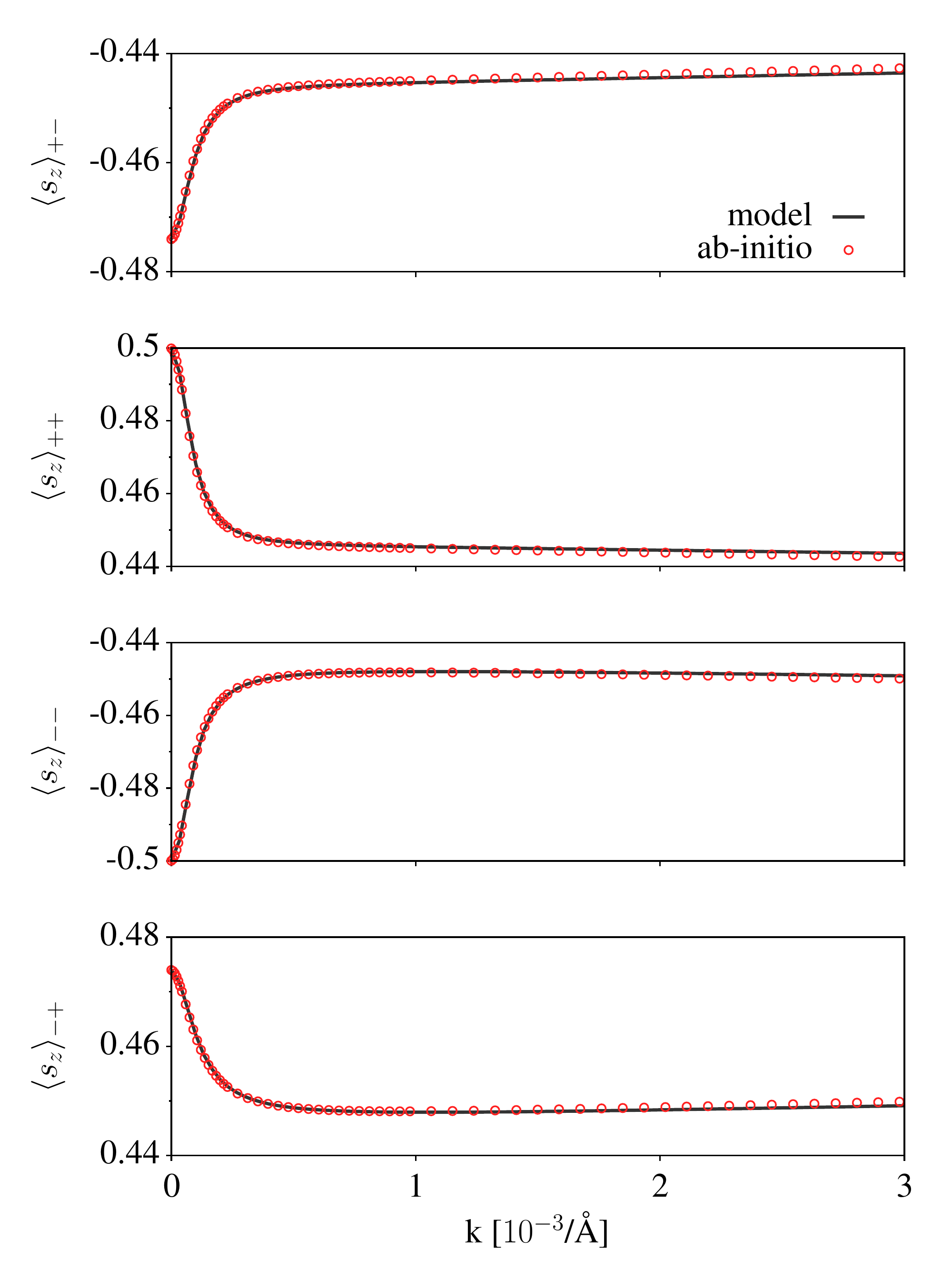}
 \caption{Calculated spin expectation values for the four low energy bands in the vicinity of the $K$ point
 for graphene on MoS$_2$. Solid lines are model calculations and symbols are first-principles data.
 }\label{Fig:sz}
\end{figure}

To obtain  $\LPIA^+$ and $\LPIA^-$  we add ${\cal H}_{\rm PIA}$ to the Hamiltonian
 ${\cal H}_0 + {\cal H}_{\rm \Delta} + {\cal H}_{\rm SO} + {\cal H}_{\rm R}$, which is used directly
at $K$ in the manuscript, and fit to the first-principles data, keeping all other parameters as determined directly at $K$. 
In Fig.~\ref{Fig:PIA}a we plot the calculated spin splittings of the valence and conduction bands.
The full model, with PIA, fits the first-principles data perfectly. The fits are 
$\LPIA^+=-4.24$~meV\AA,  and $\LPIA^-=1.25$~meV\AA. 
In Fig.~\ref{Fig:PIA}b we plot the two PIA parameters as functions of the applied transverse electric field. Their
tunability is enormous, much more than that of the Rashba hopping shown in the manuscript. 
As a final check we calculate, using the extracted parameters, the $z$ components of the spin expectation
values $\langle s_z\rangle$ in the vicinity of the $K$ point for the low energy states. The model is fully consistent
with the first-principles data, as seen in the comparison plotted in Fig.~\ref{Fig:sz}.
%


\begin{thebibliography}{43}
\expandafter\ifx\csname natexlab\endcsname\relax\def\natexlab#1{#1}\fi
\expandafter\ifx\csname bibnamefont\endcsname\relax
  \def\bibnamefont#1{#1}\fi
\expandafter\ifx\csname bibfnamefont\endcsname\relax
  \def\bibfnamefont#1{#1}\fi
\expandafter\ifx\csname citenamefont\endcsname\relax
  \def\citenamefont#1{#1}\fi
\expandafter\ifx\csname url\endcsname\relax
  \def\url#1{\texttt{#1}}\fi
\expandafter\ifx\csname urlprefix\endcsname\relax\def\urlprefix{URL }\fi
\providecommand{\bibinfo}[2]{#2}
\providecommand{\eprint}[2][]{\url{#2}}

\bibitem[{\citenamefont{Han et~al.}(2014)\citenamefont{Han, Kawakami, Gmitra,
  Fabian, Gmitra, Kawakami, and Han}}]{Han2014:NatNano}
\bibinfo{author}{\bibfnamefont{W.}~\bibnamefont{Han}},
  \bibinfo{author}{\bibfnamefont{R.~K.} \bibnamefont{Kawakami}},
  \bibinfo{author}{\bibfnamefont{M.}~\bibnamefont{Gmitra}},
  \bibinfo{author}{\bibfnamefont{J.}~\bibnamefont{Fabian}},
  \bibinfo{author}{\bibfnamefont{M.}~\bibnamefont{Gmitra}},
  \bibinfo{author}{\bibfnamefont{R.~K.} \bibnamefont{Kawakami}},
  \bibnamefont{and} \bibinfo{author}{\bibfnamefont{W.}~\bibnamefont{Han}},
  \bibinfo{journal}{Nature Nanotechnology} \textbf{\bibinfo{volume}{9}},
  \bibinfo{pages}{794} (\bibinfo{year}{2014}).

\bibitem[{\citenamefont{Tombros et~al.}(2007)\citenamefont{Tombros, J\'ozsa,
  Popinciuc, Jonkman, and van Wees}}]{Tombros2007:N}
\bibinfo{author}{\bibfnamefont{N.}~\bibnamefont{Tombros}},
  \bibinfo{author}{\bibfnamefont{C.}~\bibnamefont{J\'ozsa}},
  \bibinfo{author}{\bibfnamefont{M.}~\bibnamefont{Popinciuc}},
  \bibinfo{author}{\bibfnamefont{H.~T.} \bibnamefont{Jonkman}},
  \bibnamefont{and} \bibinfo{author}{\bibfnamefont{B.~J.} \bibnamefont{van
  Wees}}, \bibinfo{journal}{Nature} \textbf{\bibinfo{volume}{448}},
  \bibinfo{pages}{571} (\bibinfo{year}{2007}).

\bibitem[{\citenamefont{Pi et~al.}(2010)\citenamefont{Pi, Han, McCreary,
  Swartz, Li, and Kawakami}}]{Pi2010:PRL}
\bibinfo{author}{\bibfnamefont{K.}~\bibnamefont{Pi}},
  \bibinfo{author}{\bibfnamefont{W.}~\bibnamefont{Han}},
  \bibinfo{author}{\bibfnamefont{K.~M.} \bibnamefont{McCreary}},
  \bibinfo{author}{\bibfnamefont{A.~G.} \bibnamefont{Swartz}},
  \bibinfo{author}{\bibfnamefont{Y.}~\bibnamefont{Li}}, \bibnamefont{and}
  \bibinfo{author}{\bibfnamefont{R.~K.} \bibnamefont{Kawakami}},
  \bibinfo{journal}{Phys. Rev. Lett.} \textbf{\bibinfo{volume}{104}},
  \bibinfo{pages}{187201} (\bibinfo{year}{2010}).

\bibitem[{\citenamefont{Yang et~al.}(2011)\citenamefont{Yang, Balakrishnan,
  Volmer, Avsar, Jaiswal, Samm, Ali, Pachoud, Zeng, Popinciuc
  et~al.}}]{Yang2011:PRL}
\bibinfo{author}{\bibfnamefont{T.-Y.} \bibnamefont{Yang}},
  \bibinfo{author}{\bibfnamefont{J.}~\bibnamefont{Balakrishnan}},
  \bibinfo{author}{\bibfnamefont{F.}~\bibnamefont{Volmer}},
  \bibinfo{author}{\bibfnamefont{A.}~\bibnamefont{Avsar}},
  \bibinfo{author}{\bibfnamefont{M.}~\bibnamefont{Jaiswal}},
  \bibinfo{author}{\bibfnamefont{J.}~\bibnamefont{Samm}},
  \bibinfo{author}{\bibfnamefont{S.~R.} \bibnamefont{Ali}},
  \bibinfo{author}{\bibfnamefont{A.}~\bibnamefont{Pachoud}},
  \bibinfo{author}{\bibfnamefont{M.}~\bibnamefont{Zeng}},
  \bibinfo{author}{\bibfnamefont{M.}~\bibnamefont{Popinciuc}},
  \bibnamefont{et~al.}, \bibinfo{journal}{Phys. Rev. Lett.}
  \textbf{\bibinfo{volume}{107}}, \bibinfo{pages}{047206}
  (\bibinfo{year}{2011}).

\bibitem[{\citenamefont{Mak et~al.}(2010)\citenamefont{Mak, Lee, Hone, Shan,
  and Heinz}}]{Mak2010:PRL}
\bibinfo{author}{\bibfnamefont{K.~F.} \bibnamefont{Mak}},
  \bibinfo{author}{\bibfnamefont{C.}~\bibnamefont{Lee}},
  \bibinfo{author}{\bibfnamefont{J.}~\bibnamefont{Hone}},
  \bibinfo{author}{\bibfnamefont{J.}~\bibnamefont{Shan}}, \bibnamefont{and}
  \bibinfo{author}{\bibfnamefont{T.~F.} \bibnamefont{Heinz}},
  \bibinfo{journal}{Phys. Rev. Lett.} \textbf{\bibinfo{volume}{105}},
  \bibinfo{pages}{136805} (\bibinfo{year}{2010}).

\bibitem[{\citenamefont{{\v{Z}uti{\'c}}
  et~al.}(2004)\citenamefont{{\v{Z}uti{\'c}}, Fabian, and {Das
  Sarma}}}]{Zutic2004:RMP}
\bibinfo{author}{\bibfnamefont{I.}~\bibnamefont{{\v{Z}uti{\'c}}}},
  \bibinfo{author}{\bibfnamefont{J.}~\bibnamefont{Fabian}}, \bibnamefont{and}
  \bibinfo{author}{\bibfnamefont{S.}~\bibnamefont{{Das Sarma}}},
  \bibinfo{journal}{Rev. Mod. Phys.} \textbf{\bibinfo{volume}{76}},
  \bibinfo{pages}{323} (\bibinfo{year}{2004}).

\bibitem[{\citenamefont{Fabian et~al.}(2007)\citenamefont{Fabian,
  Matos-Abiague, Ertler, Stano, and \v{Z}uti\'c}}]{Fabian2007:APS}
\bibinfo{author}{\bibfnamefont{J.}~\bibnamefont{Fabian}},
  \bibinfo{author}{\bibfnamefont{A.}~\bibnamefont{Matos-Abiague}},
  \bibinfo{author}{\bibfnamefont{C.}~\bibnamefont{Ertler}},
  \bibinfo{author}{\bibfnamefont{P.}~\bibnamefont{Stano}}, \bibnamefont{and}
  \bibinfo{author}{\bibfnamefont{I.}~\bibnamefont{\v{Z}uti\'c}},
  \bibinfo{journal}{Acta Phys. Slovaca} \textbf{\bibinfo{volume}{57}},
  \bibinfo{pages}{565} (\bibinfo{year}{2007}).

\bibitem[{\citenamefont{Lopez-Sanchez et~al.}(2013)\citenamefont{Lopez-Sanchez,
  Lembke, Kayci, Radenovic, and Kis}}]{Lopez-Sanchez2013:NatNanotech}
\bibinfo{author}{\bibfnamefont{O.}~\bibnamefont{Lopez-Sanchez}},
  \bibinfo{author}{\bibfnamefont{D.}~\bibnamefont{Lembke}},
  \bibinfo{author}{\bibfnamefont{M.}~\bibnamefont{Kayci}},
  \bibinfo{author}{\bibfnamefont{A.}~\bibnamefont{Radenovic}},
  \bibnamefont{and} \bibinfo{author}{\bibfnamefont{A.}~\bibnamefont{Kis}},
  \bibinfo{journal}{Nature Nanotechnology} \textbf{\bibinfo{volume}{8}},
  \bibinfo{pages}{497} (\bibinfo{year}{2013}).

\bibitem[{\citenamefont{Huang et~al.}(2014)\citenamefont{Huang, Wu, Sanchez,
  Peters, Beanland, Ross, Rivera, Yao, Cobden, and Xu}}]{Huang2014:NatMat}
\bibinfo{author}{\bibfnamefont{C.}~\bibnamefont{Huang}},
  \bibinfo{author}{\bibfnamefont{S.}~\bibnamefont{Wu}},
  \bibinfo{author}{\bibfnamefont{A.~M.} \bibnamefont{Sanchez}},
  \bibinfo{author}{\bibfnamefont{J.~J.~P.} \bibnamefont{Peters}},
  \bibinfo{author}{\bibfnamefont{R.}~\bibnamefont{Beanland}},
  \bibinfo{author}{\bibfnamefont{J.~S.} \bibnamefont{Ross}},
  \bibinfo{author}{\bibfnamefont{P.}~\bibnamefont{Rivera}},
  \bibinfo{author}{\bibfnamefont{W.}~\bibnamefont{Yao}},
  \bibinfo{author}{\bibfnamefont{D.~H.} \bibnamefont{Cobden}},
  \bibnamefont{and} \bibinfo{author}{\bibfnamefont{X.}~\bibnamefont{Xu}},
  \bibinfo{journal}{Nature Materials} \textbf{\bibinfo{volume}{13}},
  \bibinfo{pages}{1096} (\bibinfo{year}{2014}).

\bibitem[{\citenamefont{Lee et~al.}(2014)\citenamefont{Lee, Lee, van~der Zande,
  Chen, Li, Han, Cui, Arefe, Nuckolls, Heinz et~al.}}]{Lee2014:NatNano}
\bibinfo{author}{\bibfnamefont{C.-H.} \bibnamefont{Lee}},
  \bibinfo{author}{\bibfnamefont{G.-H.} \bibnamefont{Lee}},
  \bibinfo{author}{\bibfnamefont{A.~M.} \bibnamefont{van~der Zande}},
  \bibinfo{author}{\bibfnamefont{W.}~\bibnamefont{Chen}},
  \bibinfo{author}{\bibfnamefont{Y.}~\bibnamefont{Li}},
  \bibinfo{author}{\bibfnamefont{M.}~\bibnamefont{Han}},
  \bibinfo{author}{\bibfnamefont{X.}~\bibnamefont{Cui}},
  \bibinfo{author}{\bibfnamefont{G.}~\bibnamefont{Arefe}},
  \bibinfo{author}{\bibfnamefont{C.}~\bibnamefont{Nuckolls}},
  \bibinfo{author}{\bibfnamefont{T.~F.} \bibnamefont{Heinz}},
  \bibnamefont{et~al.}, \bibinfo{journal}{Nature Nanotechnology}
  \textbf{\bibinfo{volume}{9}}, \bibinfo{pages}{676} (\bibinfo{year}{2014}).

\bibitem[{\citenamefont{Pospischil et~al.}(2014)\citenamefont{Pospischil,
  Furchi, and Mueller}}]{Pospischil2014:NatNano}
\bibinfo{author}{\bibfnamefont{A.}~\bibnamefont{Pospischil}},
  \bibinfo{author}{\bibfnamefont{M.~M.} \bibnamefont{Furchi}},
  \bibnamefont{and} \bibinfo{author}{\bibfnamefont{T.}~\bibnamefont{Mueller}},
  \bibinfo{journal}{Nature Nanotechnology} \textbf{\bibinfo{volume}{9}},
  \bibinfo{pages}{257} (\bibinfo{year}{2014}).

\bibitem[{\citenamefont{Korm{\'a}nyos et~al.}(2015)\citenamefont{Korm{\'a}nyos,
  Burkard, Gmitra, Fabian, Z{\'o}lyomi, Drummond, and
  Fal'ko}}]{Kormanyos2015:2DM}
\bibinfo{author}{\bibfnamefont{A.}~\bibnamefont{Korm{\'a}nyos}},
  \bibinfo{author}{\bibfnamefont{G.}~\bibnamefont{Burkard}},
  \bibinfo{author}{\bibfnamefont{M.}~\bibnamefont{Gmitra}},
  \bibinfo{author}{\bibfnamefont{J.}~\bibnamefont{Fabian}},
  \bibinfo{author}{\bibfnamefont{V.}~\bibnamefont{Z{\'o}lyomi}},
  \bibinfo{author}{\bibfnamefont{N.~D.} \bibnamefont{Drummond}},
  \bibnamefont{and} \bibinfo{author}{\bibfnamefont{V.}~\bibnamefont{Fal'ko}},
  \bibinfo{journal}{2D Materials} \textbf{\bibinfo{volume}{2}},
  \bibinfo{pages}{022001} (\bibinfo{year}{2015}).

\bibitem[{\citenamefont{Xiao et~al.}(2012)\citenamefont{Xiao, Liu, Feng, Xu,
  and Yao}}]{Xiao2012:PRL}
\bibinfo{author}{\bibfnamefont{D.}~\bibnamefont{Xiao}},
  \bibinfo{author}{\bibfnamefont{G.-B.} \bibnamefont{Liu}},
  \bibinfo{author}{\bibfnamefont{W.}~\bibnamefont{Feng}},
  \bibinfo{author}{\bibfnamefont{X.}~\bibnamefont{Xu}}, \bibnamefont{and}
  \bibinfo{author}{\bibfnamefont{W.}~\bibnamefont{Yao}},
  \bibinfo{journal}{Phys. Rev. Lett.} \textbf{\bibinfo{volume}{108}},
  \bibinfo{pages}{196802} (\bibinfo{year}{2012}).

\bibitem[{\citenamefont{Mak et~al.}(2012)\citenamefont{Mak, He, Shan, and
  Heinz}}]{Mak2012:NatNano}
\bibinfo{author}{\bibfnamefont{K.~F.} \bibnamefont{Mak}},
  \bibinfo{author}{\bibfnamefont{K.}~\bibnamefont{He}},
  \bibinfo{author}{\bibfnamefont{J.}~\bibnamefont{Shan}}, \bibnamefont{and}
  \bibinfo{author}{\bibfnamefont{T.~F.} \bibnamefont{Heinz}},
  \bibinfo{journal}{Nature Nanotechnology} \textbf{\bibinfo{volume}{7}},
  \bibinfo{pages}{494} (\bibinfo{year}{2012}).

\bibitem[{\citenamefont{Zeng et~al.}(2012)\citenamefont{Zeng, Dai, Yao, Xiao,
  and Cui}}]{Zeng2012:NatNano}
\bibinfo{author}{\bibfnamefont{H.}~\bibnamefont{Zeng}},
  \bibinfo{author}{\bibfnamefont{J.}~\bibnamefont{Dai}},
  \bibinfo{author}{\bibfnamefont{W.}~\bibnamefont{Yao}},
  \bibinfo{author}{\bibfnamefont{D.}~\bibnamefont{Xiao}}, \bibnamefont{and}
  \bibinfo{author}{\bibfnamefont{X.}~\bibnamefont{Cui}},
  \bibinfo{journal}{Nature Nanotechnology} \textbf{\bibinfo{volume}{7}},
  \bibinfo{pages}{490} (\bibinfo{year}{2012}).

\bibitem[{\citenamefont{Lin et~al.}(2014{\natexlab{a}})\citenamefont{Lin, Lu,
  Perea-Lopez, Li, Lin, Peng, Lee, Sun, Calderin, Browning
  et~al.}}]{Lin2014:ACS}
\bibinfo{author}{\bibfnamefont{Y.-C.} \bibnamefont{Lin}},
  \bibinfo{author}{\bibfnamefont{N.}~\bibnamefont{Lu}},
  \bibinfo{author}{\bibfnamefont{N.}~\bibnamefont{Perea-Lopez}},
  \bibinfo{author}{\bibfnamefont{J.}~\bibnamefont{Li}},
  \bibinfo{author}{\bibfnamefont{Z.}~\bibnamefont{Lin}},
  \bibinfo{author}{\bibfnamefont{X.}~\bibnamefont{Peng}},
  \bibinfo{author}{\bibfnamefont{C.~H.} \bibnamefont{Lee}},
  \bibinfo{author}{\bibfnamefont{C.}~\bibnamefont{Sun}},
  \bibinfo{author}{\bibfnamefont{L.}~\bibnamefont{Calderin}},
  \bibinfo{author}{\bibfnamefont{P.~N.} \bibnamefont{Browning}},
  \bibnamefont{et~al.}, \bibinfo{journal}{ACS Nano}
  \textbf{\bibinfo{volume}{8}}, \bibinfo{pages}{3715}
  (\bibinfo{year}{2014}{\natexlab{a}}).

\bibitem[{\citenamefont{Lin et~al.}(2014{\natexlab{b}})\citenamefont{Lin,
  Chang, Wang, Su, Chen, Lee, and Lin}}]{Lin2014:APL}
\bibinfo{author}{\bibfnamefont{M.-Y.} \bibnamefont{Lin}},
  \bibinfo{author}{\bibfnamefont{C.-E.} \bibnamefont{Chang}},
  \bibinfo{author}{\bibfnamefont{C.-H.} \bibnamefont{Wang}},
  \bibinfo{author}{\bibfnamefont{C.-F.} \bibnamefont{Su}},
  \bibinfo{author}{\bibfnamefont{C.}~\bibnamefont{Chen}},
  \bibinfo{author}{\bibfnamefont{S.-C.} \bibnamefont{Lee}}, \bibnamefont{and}
  \bibinfo{author}{\bibfnamefont{S.-Y.} \bibnamefont{Lin}},
  \bibinfo{journal}{Appl. Phys. Lett.} \textbf{\bibinfo{volume}{105}},
  \bibinfo{pages}{073501} (\bibinfo{year}{2014}{\natexlab{b}}).

\bibitem[{\citenamefont{Azizi et~al.}(2015)\citenamefont{Azizi, Eichfeld,
  Geschwind, Zhang, Jiang, Mukherjee, Hossain, Piasecki, Kabius, Robinson
  et~al.}}]{Azizi2015:ACS}
\bibinfo{author}{\bibfnamefont{A.}~\bibnamefont{Azizi}},
  \bibinfo{author}{\bibfnamefont{S.}~\bibnamefont{Eichfeld}},
  \bibinfo{author}{\bibfnamefont{G.}~\bibnamefont{Geschwind}},
  \bibinfo{author}{\bibfnamefont{K.}~\bibnamefont{Zhang}},
  \bibinfo{author}{\bibfnamefont{B.}~\bibnamefont{Jiang}},
  \bibinfo{author}{\bibfnamefont{D.}~\bibnamefont{Mukherjee}},
  \bibinfo{author}{\bibfnamefont{L.}~\bibnamefont{Hossain}},
  \bibinfo{author}{\bibfnamefont{A.~F.} \bibnamefont{Piasecki}},
  \bibinfo{author}{\bibfnamefont{B.}~\bibnamefont{Kabius}},
  \bibinfo{author}{\bibfnamefont{J.~A.} \bibnamefont{Robinson}},
  \bibnamefont{et~al.}, \bibinfo{journal}{ACS Nano}
  \textbf{\bibinfo{volume}{9}}, \bibinfo{pages}{4882} (\bibinfo{year}{2015}).

\bibitem[{\citenamefont{Lu et~al.}(2014)\citenamefont{Lu, Li, Watanabe,
  Taniguchi, and Andrei}}]{Lu2014:PRL}
\bibinfo{author}{\bibfnamefont{C.-P.} \bibnamefont{Lu}},
  \bibinfo{author}{\bibfnamefont{G.}~\bibnamefont{Li}},
  \bibinfo{author}{\bibfnamefont{K.}~\bibnamefont{Watanabe}},
  \bibinfo{author}{\bibfnamefont{T.}~\bibnamefont{Taniguchi}},
  \bibnamefont{and} \bibinfo{author}{\bibfnamefont{E.}~\bibnamefont{Andrei}},
  \bibinfo{journal}{Phys. Rev. Lett.} \textbf{\bibinfo{volume}{113}},
  \bibinfo{pages}{156804} (\bibinfo{year}{2014}).

\bibitem[{\citenamefont{Coy~Diaz et~al.}(2015)\citenamefont{Coy~Diaz, Avila,
  Chen, Addou, Asensio, and Batzill}}]{Diaz2015:NL}
\bibinfo{author}{\bibfnamefont{H.}~\bibnamefont{Coy~Diaz}},
  \bibinfo{author}{\bibfnamefont{J.}~\bibnamefont{Avila}},
  \bibinfo{author}{\bibfnamefont{C.}~\bibnamefont{Chen}},
  \bibinfo{author}{\bibfnamefont{R.}~\bibnamefont{Addou}},
  \bibinfo{author}{\bibfnamefont{M.~C.} \bibnamefont{Asensio}},
  \bibnamefont{and} \bibinfo{author}{\bibfnamefont{M.}~\bibnamefont{Batzill}},
  \bibinfo{journal}{Nano Letters} \textbf{\bibinfo{volume}{15}},
  \bibinfo{pages}{1135} (\bibinfo{year}{2015}).

\bibitem[{\citenamefont{Kumar et~al.}(2015)\citenamefont{Kumar, Dar, Gul, and
  Baek}}]{Kumar2015:MT}
\bibinfo{author}{\bibfnamefont{N.~A.} \bibnamefont{Kumar}},
  \bibinfo{author}{\bibfnamefont{M.~A.} \bibnamefont{Dar}},
  \bibinfo{author}{\bibfnamefont{R.}~\bibnamefont{Gul}}, \bibnamefont{and}
  \bibinfo{author}{\bibfnamefont{J.}~\bibnamefont{Baek}},
  \bibinfo{journal}{Materials Today} \textbf{\bibinfo{volume}{18}},
  \bibinfo{pages}{286} (\bibinfo{year}{2015}).

\bibitem[{\citenamefont{Bertolazzi et~al.}(2013)\citenamefont{Bertolazzi,
  Krasnozhon, and Kis}}]{Bertolazzi2013:ACSNano}
\bibinfo{author}{\bibfnamefont{S.}~\bibnamefont{Bertolazzi}},
  \bibinfo{author}{\bibfnamefont{D.}~\bibnamefont{Krasnozhon}},
  \bibnamefont{and} \bibinfo{author}{\bibfnamefont{A.}~\bibnamefont{Kis}},
  \bibinfo{journal}{ACS Nano} \textbf{\bibinfo{volume}{7}},
  \bibinfo{pages}{3246} (\bibinfo{year}{2013}).

\bibitem[{\citenamefont{Zhang et~al.}(2014)\citenamefont{Zhang, Chuu, Huang,
  Chen, Tsai, Chang, Liang, Chen, Chueh, He et~al.}}]{Zhang2014:SREP}
\bibinfo{author}{\bibfnamefont{W.}~\bibnamefont{Zhang}},
  \bibinfo{author}{\bibfnamefont{C.-P.} \bibnamefont{Chuu}},
  \bibinfo{author}{\bibfnamefont{J.-K.} \bibnamefont{Huang}},
  \bibinfo{author}{\bibfnamefont{C.-H.} \bibnamefont{Chen}},
  \bibinfo{author}{\bibfnamefont{M.-L.} \bibnamefont{Tsai}},
  \bibinfo{author}{\bibfnamefont{Y.-H.} \bibnamefont{Chang}},
  \bibinfo{author}{\bibfnamefont{C.-T.} \bibnamefont{Liang}},
  \bibinfo{author}{\bibfnamefont{Y.-Z.} \bibnamefont{Chen}},
  \bibinfo{author}{\bibfnamefont{Y.-L.} \bibnamefont{Chueh}},
  \bibinfo{author}{\bibfnamefont{J.-H.} \bibnamefont{He}},
  \bibnamefont{et~al.}, \bibinfo{journal}{Sci. Rep.}
  \textbf{\bibinfo{volume}{4}} (\bibinfo{year}{2014}).

\bibitem[{\citenamefont{Roy et~al.}(2013)\citenamefont{Roy, Padmanabhan,
  Goswami, Sai, Ramalingam, Raghavan, and Ghosh}}]{Roy2013:NatNanotech}
\bibinfo{author}{\bibfnamefont{K.}~\bibnamefont{Roy}},
  \bibinfo{author}{\bibfnamefont{M.}~\bibnamefont{Padmanabhan}},
  \bibinfo{author}{\bibfnamefont{S.}~\bibnamefont{Goswami}},
  \bibinfo{author}{\bibfnamefont{T.~P.} \bibnamefont{Sai}},
  \bibinfo{author}{\bibfnamefont{G.}~\bibnamefont{Ramalingam}},
  \bibinfo{author}{\bibfnamefont{S.}~\bibnamefont{Raghavan}}, \bibnamefont{and}
  \bibinfo{author}{\bibfnamefont{A.}~\bibnamefont{Ghosh}},
  \bibinfo{journal}{Nature Nanotechnology} \textbf{\bibinfo{volume}{8}},
  \bibinfo{pages}{826} (\bibinfo{year}{2013}).

\bibitem[{\citenamefont{Avsar et~al.}(2014)\citenamefont{Avsar, Tan,
  Taychatanapat, Balakrishnan, Koon, Yeo, Lahiri, Carvalho, Rodin, O’Farrell
  et~al.}}]{Avsar2014:NatComm}
\bibinfo{author}{\bibfnamefont{A.}~\bibnamefont{Avsar}},
  \bibinfo{author}{\bibfnamefont{J.~Y.} \bibnamefont{Tan}},
  \bibinfo{author}{\bibfnamefont{T.}~\bibnamefont{Taychatanapat}},
  \bibinfo{author}{\bibfnamefont{J.}~\bibnamefont{Balakrishnan}},
  \bibinfo{author}{\bibfnamefont{G.~K.~W.} \bibnamefont{Koon}},
  \bibinfo{author}{\bibfnamefont{Y.}~\bibnamefont{Yeo}},
  \bibinfo{author}{\bibfnamefont{J.}~\bibnamefont{Lahiri}},
  \bibinfo{author}{\bibfnamefont{A.}~\bibnamefont{Carvalho}},
  \bibinfo{author}{\bibfnamefont{A.~S.} \bibnamefont{Rodin}},
  \bibinfo{author}{\bibfnamefont{E.~C.~T.} \bibnamefont{O’Farrell}},
  \bibnamefont{et~al.}, \bibinfo{journal}{Nature Communications}
  \textbf{\bibinfo{volume}{5}} (\bibinfo{year}{2014}).

\bibitem[{\citenamefont{Hohenberg and Kohn}(1964)}]{Hohenberg1964:PR}
\bibinfo{author}{\bibfnamefont{J.}~\bibnamefont{Hohenberg}} \bibnamefont{and}
  \bibinfo{author}{\bibfnamefont{W.}~\bibnamefont{Kohn}},
  \bibinfo{journal}{Phys. Rev.} \textbf{\bibinfo{volume}{136}},
  \bibinfo{pages}{B864} (\bibinfo{year}{1964}).

\bibitem[{\citenamefont{Scharf and Matos-Abiague}(2012)}]{Scharf2012:PRB}
\bibinfo{author}{\bibfnamefont{B.}~\bibnamefont{Scharf}} \bibnamefont{and}
  \bibinfo{author}{\bibfnamefont{A.}~\bibnamefont{Matos-Abiague}},
  \bibinfo{journal}{Phys. Rev. B} \textbf{\bibinfo{volume}{86}},
  \bibinfo{pages}{115425} (\bibinfo{year}{2012}).

\bibitem[{\citenamefont{Gmitra et~al.}(2009)\citenamefont{Gmitra, Konschuh,
  Ertler, Ambrosch-Draxl, and Fabian}}]{Gmitra2009:PRB}
\bibinfo{author}{\bibfnamefont{M.}~\bibnamefont{Gmitra}},
  \bibinfo{author}{\bibfnamefont{S.}~\bibnamefont{Konschuh}},
  \bibinfo{author}{\bibfnamefont{C.}~\bibnamefont{Ertler}},
  \bibinfo{author}{\bibfnamefont{C.}~\bibnamefont{Ambrosch-Draxl}},
  \bibnamefont{and} \bibinfo{author}{\bibfnamefont{J.}~\bibnamefont{Fabian}},
  \bibinfo{journal}{Phys. Rev. B} \textbf{\bibinfo{volume}{80}},
  \bibinfo{pages}{235431} (\bibinfo{year}{2009}).

\bibitem[{\citenamefont{Konschuh et~al.}(2010)\citenamefont{Konschuh, Gmitra,
  and Fabian}}]{Konschuh2010:PRB}
\bibinfo{author}{\bibfnamefont{S.}~\bibnamefont{Konschuh}},
  \bibinfo{author}{\bibfnamefont{M.}~\bibnamefont{Gmitra}}, \bibnamefont{and}
  \bibinfo{author}{\bibfnamefont{J.}~\bibnamefont{Fabian}},
  \bibinfo{journal}{Phys. Rev. B} \textbf{\bibinfo{volume}{82}},
  \bibinfo{pages}{245412} (\bibinfo{year}{2010}).

\bibitem[{\citenamefont{{McClure} and Yafet}(1962)}]{j._w._mcclure_theory_1962}
\bibinfo{author}{\bibfnamefont{J.~W.} \bibnamefont{{McClure}}}
  \bibnamefont{and} \bibinfo{author}{\bibfnamefont{Y.}~\bibnamefont{Yafet}}, in
  \emph{\bibinfo{booktitle}{Proceedings of the 5th conference on carbon}}
  (\bibinfo{publisher}{Pergamon Press}, \bibinfo{year}{1962}),
  vol.~\bibinfo{volume}{1}, pp. \bibinfo{pages}{22--28}.

\bibitem[{\citenamefont{Neto and Guinea}(2009)}]{neto_impurity-induced_2009}
\bibinfo{author}{\bibfnamefont{A.~H.~C.} \bibnamefont{Neto}} \bibnamefont{and}
  \bibinfo{author}{\bibfnamefont{F.}~\bibnamefont{Guinea}},
  \bibinfo{journal}{Phys. Rev. Lett.} \textbf{\bibinfo{volume}{103}},
  \bibinfo{pages}{026804} (\bibinfo{year}{2009}).

\bibitem[{\citenamefont{Gmitra et~al.}(2013)\citenamefont{Gmitra, Kochan, and
  Fabian}}]{Gmitra2013:PRL}
\bibinfo{author}{\bibfnamefont{M.}~\bibnamefont{Gmitra}},
  \bibinfo{author}{\bibfnamefont{D.}~\bibnamefont{Kochan}}, \bibnamefont{and}
  \bibinfo{author}{\bibfnamefont{J.}~\bibnamefont{Fabian}},
  \bibinfo{journal}{Phys. Rev. Lett.} \textbf{\bibinfo{volume}{110}},
  \bibinfo{pages}{246602} (\bibinfo{year}{2013}).

\bibitem[{\citenamefont{Balakrishnan et~al.}(2013)\citenamefont{Balakrishnan,
  Kok, Koon, Jaiswal, and Neto}}]{balakrishnan_colossal_2013}
\bibinfo{author}{\bibfnamefont{J.}~\bibnamefont{Balakrishnan}},
  \bibinfo{author}{\bibfnamefont{G.}~\bibnamefont{Kok}},
  \bibinfo{author}{\bibfnamefont{W.}~\bibnamefont{Koon}},
  \bibinfo{author}{\bibfnamefont{M.}~\bibnamefont{Jaiswal}}, \bibnamefont{and}
  \bibinfo{author}{\bibfnamefont{A.~H.~C.} \bibnamefont{Neto}},
  \bibinfo{journal}{Nature Physics} \textbf{\bibinfo{volume}{9}},
  \bibinfo{pages}{1} (\bibinfo{year}{2013}).

\bibitem[{\citenamefont{Irmer et~al.}(2015)\citenamefont{Irmer, Frank, Putz,
  Gmitra, Kochan, and Fabian}}]{irmer_spin-orbit_2015}
\bibinfo{author}{\bibfnamefont{S.}~\bibnamefont{Irmer}},
  \bibinfo{author}{\bibfnamefont{T.}~\bibnamefont{Frank}},
  \bibinfo{author}{\bibfnamefont{S.}~\bibnamefont{Putz}},
  \bibinfo{author}{\bibfnamefont{M.}~\bibnamefont{Gmitra}},
  \bibinfo{author}{\bibfnamefont{D.}~\bibnamefont{Kochan}}, \bibnamefont{and}
  \bibinfo{author}{\bibfnamefont{J.}~\bibnamefont{Fabian}},
  \bibinfo{journal}{Phys. Rev. B} \textbf{\bibinfo{volume}{91}},
  \bibinfo{pages}{115141} (\bibinfo{year}{2015}).

\bibitem[{\citenamefont{Gorbachev et~al.}(2014)\citenamefont{Gorbachev, Song,
  Yu, Kretinin, Withers, Cao, Mishchenko, Grigorieva, Novoselov, Levitov
  et~al.}}]{Gorbachev2014:S}
\bibinfo{author}{\bibfnamefont{R.~V.} \bibnamefont{Gorbachev}},
  \bibinfo{author}{\bibfnamefont{J.~C.~W.} \bibnamefont{Song}},
  \bibinfo{author}{\bibfnamefont{G.~L.} \bibnamefont{Yu}},
  \bibinfo{author}{\bibfnamefont{A.~V.} \bibnamefont{Kretinin}},
  \bibinfo{author}{\bibfnamefont{F.}~\bibnamefont{Withers}},
  \bibinfo{author}{\bibfnamefont{Y.}~\bibnamefont{Cao}},
  \bibinfo{author}{\bibfnamefont{A.}~\bibnamefont{Mishchenko}},
  \bibinfo{author}{\bibfnamefont{I.~V.} \bibnamefont{Grigorieva}},
  \bibinfo{author}{\bibfnamefont{K.~S.} \bibnamefont{Novoselov}},
  \bibinfo{author}{\bibfnamefont{L.~S.} \bibnamefont{Levitov}},
  \bibnamefont{et~al.}, \bibinfo{journal}{2D Materials}
  \textbf{\bibinfo{volume}{346}}, \bibinfo{pages}{448} (\bibinfo{year}{2014}).

\bibitem[{\citenamefont{Meier and (Eds.)}(1984)}]{meier_optical_1984}
\bibinfo{author}{\bibfnamefont{F.}~\bibnamefont{Meier}} \bibnamefont{and}
  \bibinfo{author}{\bibfnamefont{B.~P.~Z.} \bibnamefont{(Eds.)}},
  \emph{\bibinfo{title}{Optical Orientation}}
  (\bibinfo{publisher}{North-Holand, New York}, \bibinfo{year}{1984}).

\bibitem[{\citenamefont{Sallen et~al.}(2012)\citenamefont{Sallen, Bouet, Marie,
  Wang, Zhu, Han, Lu, Tan, Amand, Liu et~al.}}]{Sallen2012:PRB}
\bibinfo{author}{\bibfnamefont{G.}~\bibnamefont{Sallen}},
  \bibinfo{author}{\bibfnamefont{L.}~\bibnamefont{Bouet}},
  \bibinfo{author}{\bibfnamefont{X.}~\bibnamefont{Marie}},
  \bibinfo{author}{\bibfnamefont{G.}~\bibnamefont{Wang}},
  \bibinfo{author}{\bibfnamefont{C.~R.} \bibnamefont{Zhu}},
  \bibinfo{author}{\bibfnamefont{W.~P.} \bibnamefont{Han}},
  \bibinfo{author}{\bibfnamefont{Y.}~\bibnamefont{Lu}},
  \bibinfo{author}{\bibfnamefont{P.~H.} \bibnamefont{Tan}},
  \bibinfo{author}{\bibfnamefont{T.}~\bibnamefont{Amand}},
  \bibinfo{author}{\bibfnamefont{B.~L.} \bibnamefont{Liu}},
  \bibnamefont{et~al.}, \bibinfo{journal}{Phys. Rev. B}
  \textbf{\bibinfo{volume}{86}}, \bibinfo{pages}{081301}
  (\bibinfo{year}{2012}).

\bibitem[{\citenamefont{Plechinger et~al.}(2014)\citenamefont{Plechinger,
  Nagler, Sch{\"u}ller, and Korn}}]{Korn2014:P}
\bibinfo{author}{\bibfnamefont{G.}~\bibnamefont{Plechinger}},
  \bibinfo{author}{\bibfnamefont{P.}~\bibnamefont{Nagler}},
  \bibinfo{author}{\bibnamefont{Sch{\"u}ller}}, \bibnamefont{and}
  \bibinfo{author}{\bibfnamefont{T.}~\bibnamefont{Korn}},
  \bibinfo{journal}{arXiv:1404.7674}  (\bibinfo{year}{2014}).

\bibitem[{\citenamefont{Giannozzi and et~al.}(2009)}]{Giannozzi2009:JPCM}
\bibinfo{author}{\bibfnamefont{P.}~\bibnamefont{Giannozzi}} \bibnamefont{and}
  \bibinfo{author}{\bibnamefont{et~al.}}, \bibinfo{journal}{J.Phys.: Condens.
  Matter} \textbf{\bibinfo{volume}{21}}, \bibinfo{pages}{395502}
  (\bibinfo{year}{2009}).

\bibitem[{\citenamefont{Perdew et~al.}(1996)\citenamefont{Perdew, Burke, and
  Ernzerhof}}]{Perdew1996:PRL}
\bibinfo{author}{\bibfnamefont{J.~P.} \bibnamefont{Perdew}},
  \bibinfo{author}{\bibfnamefont{K.}~\bibnamefont{Burke}}, \bibnamefont{and}
  \bibinfo{author}{\bibfnamefont{M.}~\bibnamefont{Ernzerhof}},
  \bibinfo{journal}{Phys. Rev. Lett.} \textbf{\bibinfo{volume}{77}},
  \bibinfo{pages}{3865} (\bibinfo{year}{1996}).

\bibitem[{\citenamefont{Bengtsson}(1999)}]{Bengtsson1999:PRB}
\bibinfo{author}{\bibfnamefont{L.}~\bibnamefont{Bengtsson}},
  \bibinfo{journal}{Phys. Rev. B} \textbf{\bibinfo{volume}{59}},
  \bibinfo{pages}{12301} (\bibinfo{year}{1999}).

\bibitem[{\citenamefont{Grimme}(2006)}]{Grimme2006:JCC}
\bibinfo{author}{\bibfnamefont{S.}~\bibnamefont{Grimme}}, \bibinfo{journal}{J.
  Comput. Chem.} \textbf{\bibinfo{volume}{27}}, \bibinfo{pages}{1787}
  (\bibinfo{year}{2006}).

\bibitem[{\citenamefont{Barone et~al.}(2009)\citenamefont{Barone, Casarin,
  Forrer, Pavone, Sambi, and Vittadini}}]{Barone2009:JCC}
\bibinfo{author}{\bibfnamefont{V.}~\bibnamefont{Barone}},
  \bibinfo{author}{\bibfnamefont{M.}~\bibnamefont{Casarin}},
  \bibinfo{author}{\bibfnamefont{D.}~\bibnamefont{Forrer}},
  \bibinfo{author}{\bibfnamefont{M.}~\bibnamefont{Pavone}},
  \bibinfo{author}{\bibfnamefont{M.}~\bibnamefont{Sambi}}, \bibnamefont{and}
  \bibinfo{author}{\bibnamefont{Vittadini}}, \bibinfo{journal}{J. Comput.
  Chem.} \textbf{\bibinfo{volume}{30}}, \bibinfo{pages}{934}
  (\bibinfo{year}{2009}).

\end{thebibliography}

\end{document}